\documentclass[aps,showpacs,preprint]{revtex4}
\usepackage[dvips]{graphicx}
\makeatletter
\@addtoreset{equation}{section}
\makeatother

\begin{document}
%\textsf{epsbox}
% Use the \preprint command to place your local institutional report
% number in the upper righthand corner of the title page in preprint mode.
% Multiple \preprint commands are allowed.
% Use the 'preprintnumbers' class option to override journal defaults
% to display numbers if necessary

\preprint{KOBE-TH-05-02}

%Title of paper
\title{Absolute Values of Neutrino Masses implied by the Seesaw Mechanism}

% repeat the \author .. \affiliation  \i{{\it etc.}} as needed
% \email, \thanks, \homepage, \altaffiliation all apply to the current
% author. Explanatory text should go in the []'s, actual e-mail
% address or url should go in the {}'s for \email and \homepage.
% Please use the appropriate macro foreach each type of information

% \affiliation command applies to all authors since the last
% \affiliation command. The \affiliation command should follow the
% other information
% \affiliation can be followed by \email, \homepage, \thanks as well.
\author{~H.~Tsujimoto}
\email[]{tujimoto@kobe-u.ac.jp}
\affiliation
{Department of Physics, Kobe University, Rokkodaicho 1-1, Nada ward, Kobe 657-8501, Japan}
%\homepage[]{Your web page}
%\thanks{}
%\altaffiliation{}

%Collaboration name if desired (requires use of superscriptaddress
%option in \documentclass). \noaffiliation is required (may also be
%used with the \author command).
%\collaboration can be followed by \email, \homepage, \thanks as well.
%\collaboration{}
%\noaffiliation

\date{\today}

\begin{abstract}
It is found that the seesaw mechanism not only explain the smallness of neutrino masses but also account for the large mixing angles simultaneously, even if the unification of the neutrino Dirac mass matrix with that of up-type quark sector is realized.
We show that provided the Majorana masses have hierarchical structure as is seen in the up-type quark sector and all mass matrices are real, we can reduce the information about the absolute values of neutrino masses through the data set of neutrino experiments.
Especially for $\theta _{13}=0$, we found that the neutrino masses are decided as $m_1:m_2:m_3\approx 1:3:17$ or $1:50:250$ ($m_1\simeq m_2:m_3\approx 3:1$ or $12:1$) in the case of normal mass spectrum (inverted mass spectrum), and the greatest Majorana mass turns out to be $m_3^R=1\times 10^{15}$ GeV which just corresponds to the GUT scale.
Including the decoupling effects caused by three singlet neutrinos, we also perform a renormalization group analysis to fix the neutrino Yukawa coupling matrix at low energy.
\end{abstract}

% insert suggested PACS numbers in braces on next line

% insert suggested keywords - APS authors don't need to do this

%\maketitle must follow title, authors, abstract, \pacs, and \keywords
\maketitle

% body of paper here - Use proper section commands
% References should be done using the \cite, \ref, and \label commands
%%%%%%%%%%%%%%%%%%%%%%%%%%%%%%%%%%%%%%%%%%%%%%%%%%%%%%%%%%%%%%
%%%%%%%%%%%%%%%%%%%%%%%%%%%%%%%%%%%%%%%%%%%%%%%%%%%%%%%%%%%%%%
%%%%%%%%%%%%%%%%%%%%%%%%%%%%%%%%%%%%%%%%%%%%%%%%%%%%%%%%%%%%%%
%%%%%%%%%%%%%%%%%%%%%%%%%%%%%%%%%%%%%%%%%%%%%%%%%%%%%%%%%%%%%%
%%%%%%%%%%%%%%%%%%%%%%%%%%%%%%%%%%%%%%%%%%%%%%%%%%%%%%%%%%%%%%
\section{Introduction\label{intro}}
Neutrino sector has many curious properties which are not shared by the quark and charged lepton sectors.
For example, neutrino masses are very small \cite{WMAP} compared with those of quarks and charge leptons.
The large mixing angles seen in the experiments of atmospheric neutrino oscillation and long baseline reactor neutrino oscillation (related to solar neutrino deficit) \cite{SK,Kam1,Kam2,SNO} are also the new feature, not seen in the quark sector.

It is well known that the seesaw mechanism \cite{Yanagida,Gell-Mann,Mohapatra} can explain the smallness of neutrino masses naturally.
In this mechanism, neutrino mass matrix which describes the low energy observables is given approximately by
\begin{eqnarray}
\mathcal{M}_{\nu}
&=&
-\mathcal{M}_D^T\mathcal{M}_R^{-1}\mathcal{M}_D,
\label{nmass1}
\end{eqnarray}
where $\mathcal{M}_D$ and $\mathcal{M}_R$ are Dirac and Majorana mass matrices of neutrino, respectively.
If we require that the order of magnitude of $\mathcal{M}_D$ is the weak scale and that of $\mathcal{M}_R$ is the GUT scale, we can roughly obtain the desired order of magnitude of $\mathcal{M}_{\nu}$.

In addition, it was pointed out in some articles \cite{Smirnov1} that this mechanism also may be responsible for the enhancement of the mixings in the leptonic sector, compared with those in quark sector.
This enhancement mechanism can be seen in the case of simplified two-generation scheme as follows.
 If we assume that $\mathcal{M}_D$ is symmetric and real matrix and $\mathcal{M}_R$ is real matrix (and necessarily symmetric), we can represent $\mathcal{M}_D$ and $\mathcal{M}_R^{-1}$ as
\begin{eqnarray}
\mathcal{M}_D
&=&
\left[
\begin{array}{cc}
c_{\theta}&s_{\theta}\\
-s_{\theta}&c_{\theta}\\
\end{array}
\right]
\left[
\begin{array}{cc}
m_1^D&0\\
0&m_2^D\\
\end{array}
\right]
\left[
\begin{array}{cc}
c_{\theta}&-s_{\theta}\\
s_{\theta}&c_{\theta}\\
\end{array}
\right]
\nonumber\\
&=&
m_2^D\cdot
\left[
\begin{array}{cc}
c_{\theta}&s_{\theta}\\
-s_{\theta}&c_{\theta}\\
\end{array}
\right]
\left[
\begin{array}{cc}
r&0\\
0&1\\
\end{array}
\right]
\left[
\begin{array}{cc}
c_{\theta}&-s_{\theta}\\
s_{\theta}&c_{\theta}\\
\end{array}
\right]
\\
\mathcal{M}_R^{-1}
&=&
\left[
\begin{array}{cc}
c_{\phi}&s_{\phi}\\
-s_{\phi}&c_{\phi}\\
\end{array}
\right]
\left[
\begin{array}{cc}
\frac{1}{m_1^R}&0\\
0&\frac{1}{m_2^R}\\
\end{array}
\right]
\left[
\begin{array}{cc}
c_{\phi}&-s_{\phi}\\
s_{\phi}&c_{\phi}\\
\end{array}
\right]
\nonumber\\
&=&
\frac{1}{m_1^R}
\left[
\begin{array}{cc}
c_{\phi}&s_{\phi}\\
-s_{\phi}&c_{\phi}\\
\end{array}
\right]
\left[
\begin{array}{cc}
1&0\\
0&R\\
\end{array}
\right]
\left[
\begin{array}{cc}
c_{\phi}&-s_{\phi}\\
s_{\phi}&c_{\phi}\\
\end{array}
\right]
,
\label{}
\end{eqnarray}
where $r=m_1^D/m_2^D,~R=m_1^R/m_2^R$ and $c_{\theta}=\cos\theta,~s_{\theta}=\sin\theta$ {\it etc.}
We have assumed that $\mathcal{M}_D$ coincides with that of up-type quarks in the basis where down-type quarks are in their mass eigenstate.
Namely $\theta$ is the Cabibbo angle.
$\mathcal{M}_{\nu}$ is then represented as
\begin{eqnarray}
\mathcal{M}_{\nu}
&=&
-\frac{{m_2^D}^2}{m_1^R}
\left[
\begin{array}{cc}
c_{\theta}&s_{\theta}\\
-s_{\theta}&c_{\theta}\\
\end{array}
\right]
\left[
\begin{array}{cc}
r^2\left(c_{\theta-\phi}^2+Rs_{\theta-\phi }^2\right)&
r\left(1-R\right)s_{\theta-\phi }c_{\theta-\phi }\\
r\left(1-R\right)s_{\theta-\phi }c_{\theta-\phi }&
s_{\theta-\phi }^2+Rc_{\theta-\phi }^2\\
\end{array}
\right]
\left[
\begin{array}{cc}
c_{\theta}&-s_{\theta}\\
s_{\theta}&c_{\theta}\\
\end{array}
\right]
\nonumber\\
&=&
-
\left[
\begin{array}{cc}
c_{\theta}&s_{\theta}\\
-s_{\theta}&c_{\theta}\\
\end{array}
\right]
\left[
\begin{array}{cc}
c_{\alpha}&s_{\alpha}\\
-s_{\alpha}&c_{\alpha}\\
\end{array}
\right]
\left[
\begin{array}{cc}
m_1&0\\
0&m_2\\
\end{array}
\right]
\left[
\begin{array}{cc}
c_{\alpha}&-s_{\alpha}\\
s_{\alpha}&c_{\alpha}\\
\end{array}
\right]
\left[
\begin{array}{cc}
c_{\theta}&-s_{\theta}\\
s_{\theta}&c_{\theta}\\
\end{array}
\right]
\nonumber\\
&=&
-
\left[
\begin{array}{cc}
c_{\theta+\alpha}&s_{\theta+\alpha}\\
-s_{\theta+\alpha}&c_{\theta+\alpha}\\
\end{array}
\right]
\left[
\begin{array}{cc}
m_1&0\\
0&m_2\\
\end{array}
\right]
\left[
\begin{array}{cc}
c_{\theta+\alpha}&-s_{\theta+\alpha}\\
s_{\theta+\alpha}&c_{\theta+\alpha}\\
\end{array}
\right],
\label{}
\end{eqnarray}
where
\begin{eqnarray}
\tan 2\alpha
&=&
\frac{2r(1-R)\tan(\theta-\phi)}{R-r^2+(1-r^2R)\tan^2(\theta-\phi)}
\nonumber\\
&\simeq&
\frac{2r\tan(\theta-\phi)}{R-r^2+\tan^2(\theta-\phi)}.
\label{}
\end{eqnarray}
This means that in the leptonic sector the mixing angle is replaced as $\theta\rightarrow\theta+\alpha$ and if the condition
\begin{eqnarray}
R-r^2+\tan^2(\theta-\phi)\ll r\tan(\theta-\phi)
\label{}
\end{eqnarray}
is satisfied, $\alpha$ becomes large considerably and the enhancement of mixing angle is realized due to the seesaw mechanism.
In this context, the large mixing angle $\theta+\alpha$ is induced only for certain relation between three parameters, i.e. $r,R,\tan(\theta-\phi)$.

Inspired by this observation, in what follows, we apply this mechanism to the case of realistic three-generation scheme, assuming that $\mathcal{M}_D$ is a symmetric matrix and all mass matrices are real.
The former assumption is justified in the case of $SO(10)$ GUT, because this model contains a subgroup $SU(2)_L\times SU(2)_R\times U(1)_{B-L}$ and has left-right symmetry.
Since the exchange of chirality $L\leftrightarrow R$ corresponds to $\mathcal{M}_D\rightarrow \mathcal{M}_D^T$, the left-right symmetry implies that $\mathcal{M}_D$ is a symmetric matrix, if we neglect the radiative corrections completely.
The framework of $SO(10)$ GUT, we work on, will be discussed in some detail in Sec.\ref{rg}.
The l.h.s. of Eq.(\ref{nmass1}), being a real symmetric matrix, has 6 degrees of freedom which are related to the low energy observables, i.e. 3 mixing angles and 3 mass eigenvalues.
On the other hand, the r.h.s. of eq(\ref{nmass1}) is parametrized by 12 parameters, i.e. 6 (3+3) mixing angles and 6 (3+3) mass eigenvalues of Dirac and Majorana mass matrices.
These mean that the observables are not enough to determine the parameters.
If, however, we regard $\mathcal{M}_D$ as that of up-type quark sector, as implied by $SO(10)$ GUT, and assume that the mass ratios of Majorana masses have a hierarchical structure in analogy to what we observe in quark sector, i.e.
\begin{eqnarray}
m_1^R:m_2^R:m_3^R&=&R^2:R:1,
\label{assume2}
\end{eqnarray}
the number of parameters is reduced to 5, so that the 5 parameters can be fixed by 5 observables in neutrino oscillation.
Namely, if we use the following 5 observables \cite{global},
\begin{eqnarray}
\tan^2\theta _{sol}&=&0.39_{-0.04}^{+0.05}\label{t12}\\
\tan^2\theta _{atm}&=&1_{-0.26}^{+0.35}\label{t23}\\
\sin^2\theta _{13}&\le&0.041\label{t13}\\
\Delta m_{sol}^2\approx m_2^2-m_1^2&=&\left[8.2_{-0.3}^{+0.3}\right]\times 10^{-5}~\textrm{eV}^2\label{msol}\\
\Delta m_{atm}^2\approx |m_3^2-m_2^2|&=&\left[2.2_{-0.4}^{+0.6}\right]\times 10^{-3}~\textrm{eV}^2\label{matm},
\end{eqnarray}
the 5 parameters are fixed and we can finally estimate a remaining observable, i.e. the absolute value of neutrino mass.
Identification of $\mathcal{M}_D$ with that of up-type quark sector is justified only at the GUT scale, not at low energies.
We perform a renormalization group (RG) improved analysis in Sec.\ref{rg}.

This paper is organized as follows.
In Sec.\ref{three}, we parametrize the mass matrices in the realistic three-generation scheme, and discuss how the mixing parameters out of 5 parameters are describable in terms of $R$ with the help of neutrino oscillation data.
In Sec.\ref{ana}, we utilize the remaining datum of neutrino oscillation to determine $R$ and finally derive the absolute values of neutrino masses.
In Sec.\ref{rg}, we improve these estimations taking into account the renormalization group (RG) effects to $\mathcal{M}_D$ including the effect of successive decoupling of three heavy singlet neutrinos.
Sec.\ref{concl} is devoted to the conclusion.
%%%%%%%%%%%%%%%%%%%%%%%%%%%%%%%%%%%%%%%%%%%%%%%%%%%%%%%%%%%%%%
%%%%%%%%%%%%%%%%%%%%%%%%%%%%%%%%%%%%%%%%%%%%%%%%%%%%%%%%%%%%%%
%%%%%%%%%%%%%%%%%%%%%%%%%%%%%%%%%%%%%%%%%%%%%%%%%%%%%%%%%%%%%%
%%%%%%%%%%%%%%%%%%%%%%%%%%%%%%%%%%%%%%%%%%%%%%%%%%%%%%%%%%%%%%
%%%%%%%%%%%%%%%%%%%%%%%%%%%%%%%%%%%%%%%%%%%%%%%%%%%%%%%%%%%%%%
\section{the three-generation scheme\label{three}}
In this section, we discuss the realistic three-generation scheme.
We adopt the following minimal set of assumptions:
\begin{itemize}
\item the Dirac mass matrix, $\mathcal{M}_D$, is symmetric and real.
\item the Majorana mass matrix, $\mathcal{M}_R$, is real.
\item $\mathcal{M}_D$ coincides with the mass matrix of up-type quark sector at the weak scale.
\item the eigenvalues of Majorana mass matrix have a hierarchical structure and are well approximated by $m_1^R:m_2^R:m_3^R=R^2:R:1$.
\end{itemize}

In what follows, we adopt the basis in which down-type quarks and charged leptons are in their mass eigenstates, as suggested by $SO(10)$ GUT, and use the following values for the Dirac mass matrix of neutrino at the weak scale (the 3rd assumption) \cite{Koide} ;
\begin{eqnarray}
m_1^D=2.33~\textrm{MeV}
\qquad
m_2^D=677~\textrm{MeV}
\qquad
m_3^D=181~\textrm{GeV}
\\
\sin\theta _{12}^D=0.2205
\qquad
\sin\theta _{23}^D=0.0373
\qquad
\sin\theta _{13}^D=0.003
\label{}
\end{eqnarray}
%%%%%%%%%%%%%%%%%%%%%%%%%%%%%%%%%%%%%%%%%%%%%%%%%%%%%%%%%%%%%%
%%%%%%%%%%%%%%%%%%%%%%%%%%%%%%%%%%%%%%%%%%%%%%%%%%%%%%%%%%%%%%
\subsection{Parametrization\label{para}}
According to the above assumptions, Dirac mass matrix is parametrized as
\begin{eqnarray}
\mathcal{M}_D
&=&
v
\left[
\begin{array}{ccc}
m_1^D&&\\
&m_2^D&\\
&&m_3^D\\
\end{array}
\right]
v^T
\nonumber\\
&=&
m_3^D\cdot
v
\left[
\begin{array}{ccc}
r^2&&\\
&r&\\
&&1\\
\end{array}
\right]
v^T,
\label{diracmass}
\end{eqnarray}
where we use an approximation that the Dirac neutrino masses are represented as
\begin{eqnarray}
m_1^D:m_2^D:m_3^D\approx (1/290)^2:1/290:1\equiv r^2:r:1.
\label{}
\end{eqnarray}
As an explicit representation of $v$, we use a standard parametrization \cite{PDG,Chau,Harari,Fritzsch,Botella}
\begin{eqnarray}
v^T
&=&
v_{23}v_{13}v_{12}
\nonumber\\
&=&
\left[
\begin{array}{ccc}
1&0&0\\
0&\cos\theta _{23}^D&\sin\theta _{23}^D\\
0&-\sin\theta _{23}^D&\cos\theta _{23}^D\\
\end{array}
\right]
\left[
\begin{array}{ccc}
\cos\theta _{13}^D&0&\sin\theta _{13}^D\\
0&1&0\\
-\sin\theta _{13}^D&0&\cos\theta _{13}^D\\
\end{array}
\right]
\left[
\begin{array}{ccc}
\cos\theta _{12}^D&\sin\theta _{12}^D&0\\
-\sin\theta _{12}^D&\cos\theta _{12}^D&0\\
0&0&1\\
\end{array}
\right]
\nonumber\\
&=&
\left[
\begin{array}{ccc}
c_{12}c_{13}&s_{12}c_{13}&s_{13}\\
-s_{12}c_{23}-c_{12}s_{23}s_{13}&c_{12}c_{23}-s_{12}s_{23}s_{13}&s_{23}c_{13}\\
s_{12}s_{23}-c_{12}c_{23}s_{13}&-c_{12}s_{23}-s_{12}c_{23}s_{13}&c_{23}c_{13}\\
\end{array}
\right],
\label{diracmix}
\end{eqnarray}
where $c_{12}$ stands for $\cos\theta _{12}^D$ {\it etc}.

On the other hand, Majorana mass matrix is represented as
\begin{eqnarray}
\mathcal{M}_R^{-1}
&=&
u
\left[
\begin{array}{ccc}
\frac{1}{m_1^R}&&\\
&\frac{1}{m_2^R}&\\
&&\frac{1}{m_3^R}\\
\end{array}
\right]
u^T
\nonumber\\
&=&
\frac{1}{m_1^R}\cdot
u
\left[
\begin{array}{ccc}
1&&\\
&R&\\
&&R^2\\
\end{array}
\right]
u^T,
\label{mayoranamass}
\end{eqnarray}
where $u$ is an orthogonal matrix.

Combining Eq.(\ref{diracmass}) and Eq.(\ref{mayoranamass}), we can represent neutrino mass matrix as
\begin{eqnarray}
\mathcal{M}_{\nu}
&=&
-
\mathcal{M}_D^T\mathcal{M}_R^{-1}\mathcal{M}_D
\nonumber\\
&=&
-
\frac{(m_3^D)^2}{m_1^R}\cdot
v
\left[
\begin{array}{ccc}
r^2&&\\
&r&\\
&&1\\
\end{array}
\right]
V^T
\left[
\begin{array}{ccc}
1&&\\
&R&\\
&&R^2\\
\end{array}
\right]
V
\left[
\begin{array}{ccc}
r^2&&\\
&r&\\
&&1\\
\end{array}
\right]
v^T.
\label{neutrinomass}
\end{eqnarray}
where $V\equiv u^Tv$ denotes the deviation of $u$ from CKM matrix, and contains 3 mixing angles.
On the other hand, l.h.s of Eq.(\ref{neutrinomass}) can be written by observables as
\begin{eqnarray}
\mathcal{M}_{\nu}
&=&
U
\left[
\begin{array}{ccc}
m_1&&\\
&m_2&\\
&&m_3\\
\end{array}
\right]
U^T,
\label{neutrinomassobs}
\end{eqnarray}
where $m_i$'s are neutrino masses and $U$ is the MNS matrix except for the CP phase,
\begin{eqnarray}
U
&\equiv&
U_{23}U_{13}U_{12}
\nonumber\\
&=&
\left[
\begin{array}{ccc}
1&0&0\\
0&\cos\theta _{23}&\sin\theta _{23}\\
0&-\sin\theta _{23}&\cos\theta _{23}\\
\end{array}
\right]
\left[
\begin{array}{ccc}
\cos\theta _{13}&0&\sin\theta _{13}\\
0&1&0\\
-\sin\theta _{13}&0&\cos\theta _{13}\\
\end{array}
\right]
\left[
\begin{array}{ccc}
\cos\theta _{12}&\sin\theta _{12}&0\\
-\sin\theta _{12}&\cos\theta _{12}&0\\
0&0&1\\
\end{array}
\right].
\label{mnsmatrix}
\end{eqnarray}
These observables have already got certain values or constraints as seen in Eqs.(\ref{t12})(\ref{t23})(\ref{t13})(\ref{msol})(\ref{matm}).

For a while we concentrate on the mixing angles and mass ratios, ignoring $(m_3^D)^2/m_1^R$ in Eq.(\ref{neutrinomass}).
Then the following 4 observables obtained from the neutrino oscillation data
\begin{eqnarray}
\tan^2\theta _{12}&=&0.39\label{t12rat}\\
\tan^2\theta _{23}&=&1\label{t23rat}\\
\sin^2\theta _{13}&=&0\label{t13rat}\\
\frac{m_2^2-m_1^2}{|m_3^2-m_2^2|}&=&\frac{8.2\times 10^{-5}}{2.2\times 10^{-3}}\label{massrat},
\end{eqnarray}
are enough to fix the (3 angles in the) mixing matrix $V$ and the mass ratio $R$.
%%%%%%%%%%%%%%%%%%%%%%%%%%%%%%%%%%%%%%%%%%%%%%%%%%%%%%%%%%%%%%
%%%%%%%%%%%%%%%%%%%%%%%%%%%%%%%%%%%%%%%%%%%%%%%%%%%%%%%%%%%%%%
\subsection{Constraints to mixing matrix\label{const}}
To decide the mixing matrix $V$ and $R$, we define a matrix $A$ by taking off a factor $-(m_3^D)^2/m_1^R$ from $\mathcal{M}_{\nu}$ in Eq.(\ref{neutrinomass})
\begin{eqnarray}
A
&\equiv&
v
\left[
\begin{array}{ccc}
r^2&&\\
&r&\\
&&1\\
\end{array}
\right]
V^T
\left[
\begin{array}{ccc}
1&&\\
&R&\\
&&R^2\\
\end{array}
\right]
V
\left[
\begin{array}{ccc}
r^2&&\\
&r&\\
&&1\\
\end{array}
\right]
v^T
\label{Adef}\\
&\equiv&
\left[
\begin{array}{ccc}
A_{11}&A_{12}&A_{13}\\
A_{12}&A_{22}&A_{23}\\
A_{13}&A_{23}&A_{33}\\
\end{array}
\right].
\end{eqnarray}
As this matrix is proportional to $\mathcal{M}_{\nu}$, it should be diagonalized by MNS matrix $U$.
Namely,
\begin{eqnarray}
A
&\propto&
U
\left[
\begin{array}{ccc}
\frac{m_1}{m_3}&&\\
&\frac{m_2}{m_3}&\\
&&1\\
\end{array}
\right]
U^T.
\label{apropto}
\end{eqnarray}
Then by using Eq.(\ref{t12rat}) to Eq.(\ref{t13rat}), i.e. $\theta _{23}=45^{\circ}, \theta _{13}=0^{\circ}$ and $\theta _{12}\equiv\theta=32^{\circ}$ (this $\theta$ should not be confused with that used in the two-generation scheme), for the mixing in Eq.(\ref{mnsmatrix}), we get from this relation 3 conditions to be satisfied by the elements of $A$:
\begin{eqnarray}
A_{22}=A_{33}\label{t23cond}\\
A_{12}=-A_{13}\label{t13cond}\\
\frac{2\sqrt{2}A_{12}}{A_{22}-A_{23}-A_{11}}=\tan 2\theta\label{t12cond}.
\end{eqnarray}
We also have relations to express mass ratios, $\rho _1,\rho _2$, in terms of the elements of $A$:
\begin{eqnarray}
\rho _1
&\equiv&
\frac{m_1}{m_3}
\nonumber\\
&=&
\frac{A_{22}-A_{23}-\sqrt{2}\cot\theta A_{12}}{A_{22}+A_{23}},
\label{rho1}
\\
\rho _2
&\equiv&
\frac{m_2}{m_3}
\nonumber\\
&=&
\frac{A_{22}-A_{23}+\sqrt{2}\tan\theta A_{12}}{A_{22}+A_{23}}.
\label{rho2}
\end{eqnarray}

Basically, the 3 conditions in Eqs.(\ref{t23cond}),(\ref{t13cond}),(\ref{t12cond}) can be used to express the 3 mixing angles in $V$, i.e. $\phi _{23},\phi _{13},\phi _{12}$ in $V\equiv V_{23}V_{13}V_{12}$, in terms of $R$.
Then the matrix $A$ is completely written in terms of $R$, and substituting Eqs.(\ref{rho1}),(\ref{rho2}) in Eq.(\ref{massrat}), $R$ can be fixed.
Using Eq.(\ref{msol})(\ref{matm}), we can finally determine all absolute values of neutrino masses, $m_1,m_2$ and $m_3$.
The matrix $A$, however, has a complicated form when written in terms of $\phi _{12},\phi _{23},\phi _{13}$ and $R$, so that it is not easy to represent $V$ in terms of $R$.
To avoid this difficulty, we will utilize the orthonormality of $V$ as much as possible, and will use only a minimal approximation.
%%%%%%%%%%%%%%%%%%%%%%%%%%%%%%%%%%%%%%%%%%%%%%%%%%%%%%%%%%%%%%
%%%%%%%%%%%%%%%%%%%%%%%%%%%%%%%%%%%%%%%%%%%%%%%%%%%%%%%%%%%%%%
\subsection{Appropriate approximation\label{approx}}
It is worth noting that, because of the orthonormality of $V$, for $R=1$ the matrix $V$ becomes irrelevant, since
\begin{eqnarray}
V^T\Lambda V=V^T(I-(I-\Lambda))V=I-V^T(I-\Lambda)V,
\label{}
\end{eqnarray}
where $\Lambda\equiv diag(1,R,R^2)$.
To achieve a minimal approximation mentioned above, let us define a matrix $X$ as
\begin{eqnarray}
X\equiv V^T(I-\Lambda)V,
\label{xdef}
\end{eqnarray}
and discuss about the relations between $X$ and $R$, instead of discussing $\phi _{12},\phi _{23},\phi _{13}$ themselves.
As $X$ is a symmetric matrix described by $V_{2i},V_{3i}~(i=1-3)$, it contains 6 components, which we denote as $X_1$ to $X_6$: $X_1\equiv X_{11},X_2\equiv X_{12},X_3\equiv X_{13},X_4\equiv X_{22},X_5\equiv X_{23},X_6\equiv X_{33}$.
At first glance, one may think that the degrees of freedom in both sides of Eq.(\ref{xdef}) don't coincide.
We, however, can avoid this mismatch utilizing the orthonormality of $V$ as is seen below. 

Using above definition and Eq.(\ref{Adef}), each element of $A$ may be written as
\begin{eqnarray}
A_{ij}
&=&
a_{ij}^{(0)}
-\sum _{k=1}^{6}a_{ij}^{(k)}X_k,
\label{Aelements}
\end{eqnarray}
where $a_{ij}$'s are defined as
\begin{eqnarray}
a_{ij}^{(0)}
&\equiv&
\sum _{k=1}^{3}r^{6-2k}\cdot v_{ik}v_{jk}
\label{a0}\\
a_{ij}^{(1)}
&\equiv&
r^4\cdot v_{i1}v_{j1}
\label{a1}\\
a_{ij}^{(2)}
&\equiv&
r^3\cdot (v_{i1}v_{j2}+v_{i2}v_{j1})
\label{a2}\\
a_{ij}^{(3)}
&\equiv&
r^2\cdot (v_{i1}v_{j3}+v_{i3}v_{j1})
\label{a3}\\
a_{ij}^{(4)}
&\equiv&
r^2\cdot v_{i2}v_{j2}
\label{a4}\\
a_{ij}^{(5)}
&\equiv&
r\cdot (v_{i2}v_{j3}+v_{i3}v_{j2})
\label{a5}\\
a_{ij}^{(6)}
&\equiv&
v_{i3}v_{j3}
\label{a6}
\end{eqnarray}

Then the three conditions, Eqs.(\ref{t23cond}),(\ref{t13cond}),(\ref{t12cond}), lead to coupled linear equations for $X_i$:
\begin{eqnarray}
\bullet~A_{22}=A_{33}
&\Rightarrow&
\sum _{k=1}^{6}(a_{22}^{(k)}-a_{33}^{(k)})X_k=a^{(0)}_{22}-a_{33}^{(0)}\label{cond1}\\
\bullet~A_{12}=-A_{13}
&\Rightarrow&
\sum _{k=1}^{6}(a_{12}^{(k)}+a_{13}^{(k)})X_k=a^{(0)}_{12}+a_{13}^{(0)}\label{cond2}\\
\bullet~\frac{2\sqrt{2}A_{12}}{A_{22}-A_{23}-A_{11}}=\tan 2\theta
&\Rightarrow&
\sum _{k=1}^{6}(a_{11}^{(k)}-a_{22}^{(k)}+a_{23}^{(k)}+2\sqrt{2}\cot 2\theta a_{12}^{(k)})X_k\nonumber\\
&&=a^{(0)}_{11}-a^{(0)}_{22}+a_{23}^{(0)}+2\sqrt{2}\cot 2\theta a_{12}^{(0)}\label{cond3}.
\end{eqnarray}

As mentioned above, not all of $X_i$ are independent by orthonormality.
For instance, due to the orthonormality of $V$ we find $\textrm{Tr(X)}=\textrm{Tr}(I-\Lambda)=1-R+1-R^2$, i.e.
\begin{eqnarray}
X_1+X_4+X_6&=&1-R+1-R^2.
\label{cond4}
\end{eqnarray}
Similarly $\textrm{Tr}(X^2)$ yields
\begin{eqnarray}
X_1^2+2X_2^2+2X_3^2+X_4^2+2X_5^2+X_6^2&=&(1-R)^2+(1-R^2)^2.
\label{cond5}
\end{eqnarray}
Another relation comes from the determinant of $X$:
\begin{eqnarray}
X_1X_4X_6-X_1X_5^2-X_4X_3^2-X_6X_2^2+2X_2X_3X_5=0.
\label{cond8}
\end{eqnarray}
In this way, we have found, in total, 6 coupled equations for $X_i$, which are enough to solve for $X _i$ in terms of $R$.
However, two of them are non-linear equations and are very difficult to solve in terms of $R$ analytically
\footnote{In general, it is also not so easy to solve non-linear coupled equations numerically.}.
Alternatively, we will make one approximation in Eq.(\ref{cond4}) and divide it into two equations.
Using Eq.(\ref{xdef}), we can express $X_1+X_4$ as
\begin{eqnarray}
X_1+X_4
=
(1-R)\left\{1+(1+R)V_{13}^2+RV_{23}^2\right\}.
\nonumber
\end{eqnarray}
Assuming that $V_{13}\ll 1$ and $R\ll 1$, we use following approximated expressions.
\begin{eqnarray}
X_1+X_4&\simeq&1-R\label{cond6}\\
X_6&\simeq&1-R^2\label{cond7}
\label{}
\end{eqnarray}
When these equations are combined with the linear three conditions (\ref{cond1}),(\ref{cond2}),(\ref{cond3}), we can represent $X_2,X_3,X_4,X_5$ and $X_6$ in terms of linear combinations of $X_1$ and $R$.
Substituting these in the remaining non-linear equation (\ref{cond5}), we obtain a quadratic equation about $X_1$ which can be solved easily and finally obtain the expressions for $X_1,\cdots,X_6$ as functions of $R$.
We can finally fix the value of $R$ by utilizing the remaining condition (\ref{massrat}), i.e.
\begin{eqnarray}
\frac{\rho _2^2(R)-\rho _1^2(R)}{|1-\rho _2^2(R)|}
=
\frac{8.2\times 10^{-5}}{2.2\times 10^{-3}}.
\label{lastcond}
\end{eqnarray}
Note that there are two possible cases reflecting the uncertainty of the sign of mass-squared difference in atmospheric neutrino oscillation experiment, i.e. normal or inverted mass spectrum, i.e. $\Delta m_{atm}^2>0$ or $\Delta m_{atm}^2<0$, respectively. 
There are some proposals to fix the sign of atmospheric neutrino mass squared difference, i.e. discrimination between normal mass spectrum and inverted one by utilizing the difference of matter effect of the earth between electron neutrino and electron anti-neutrino at Neutrino Factory\cite{Albright,Barger}.
%%%%%%%%%%%%%%%%%%%%%%%%%%%%%%%%%%%%%%%%%%%%%%%%%%%%%%%%%%%%%%
%%%%%%%%%%%%%%%%%%%%%%%%%%%%%%%%%%%%%%%%%%%%%%%%%%%%%%%%%%%%%%
\section{Numerical results\label{ana}}
%%%%%%%%%%%%%%%%%%%%%%%%%%%%%%%%%%%%%%%%%%%%%%%%%%%%%%%%%%%%%%
\subsection{Normal mass spectrum\label{nmh}}
In the case of normal mass spectrum, two values of $R$ are allowed
\begin{eqnarray}
R
=
(3.214,-3.780)\times 10^{-5}
=
(2.703,-3.179)\times r^2.
\label{Rf}
\end{eqnarray}
Substituting these values in Eqs.(\ref{rho1}),(\ref{rho2}), we can immediately calculate the values of $\rho _1$ and $\rho _2$, i.e. the ratios of neutrino masses
\begin{eqnarray}
\rho _1
&\equiv&
\frac{m_1}{m_3}
=
(-0.003977,-0.05856),
\label{rho1Rf}\\
\rho _2
&\equiv&
\frac{m_2}{m_3}
=
(0.1896,0.1981).
\label{rho2Rf}
\end{eqnarray}
We can interpret the sign of mass eigenvalues seen in Eqs.(\ref{rho1Rf}),(\ref{rho2Rf}) including the overall one in Eq.(\ref{neutrinomass}) as CP-parity in neutrino sector and can define all of mass eigenvalues as positive without a conflict with CP-invariance.
For detailed discussion about CP-invariance and CP-parity, see refs.\cite{Bilenky,Bilenky2}.

Taking into account this fact, we can eventually determine the absolute values of neutrino masses, invoking to the best fit values of Eqs.(\ref{msol}),(\ref{matm}) :
\begin{eqnarray}
m_1
&=&
(0.1900,2.802)\times 10^{-3}~\textrm{eV}
\label{m1f}\\
m_2
&=&
(9.057,9.479)\times 10^{-3}~\textrm{eV}
\label{m2f}\\
m_3
&=&
(47.77,47.85)\times 10^{-3}~\textrm{eV}.
\label{m3f}
\end{eqnarray}
Furthermore, from Eqs.(\ref{neutrinomass}),(\ref{neutrinomassobs}) and $m_3^D=181$ GeV, we can estimate the greatest Majorana mass as
\begin{eqnarray}
m_3^R
&=&
\frac{(m_3^D)^2}{m_3}\cdot\frac{\left|A_{22}+A_{23}\right|}{R^2}
=
(1.431,1.401)\times 10^{15}~\textrm{GeV}.
\label{majoranamassf}
\end{eqnarray}

Substituting the obtained values of $R$ to Eq.(\ref{cond8}), we can immediately calculate the values of $det(X)$ and we can find that it does not vanish exactly but has an extremely small value, $\mathcal{O}(10^{-13})$, which can be regarded as the error of the numerical calculation in the last step to fix $R$.
We also can confirm that the approximate relation, (\ref{cond6}),(\ref{cond7}), hold within this error.

It is worth mentioning that the absolute values of neutrino masses at low energy obtained above are similar to the results obtained in Ref.\cite{Mahan}, in which the analysis are based on SUSY $SO(10)\times SU(2)_F$.
They make use of quark-lepton symmetry and left-right symmetry at $M_{GUT}$ relying on $SO(10)$, and a specific mass matrices with five texture zeros due to $SU(2)_F$ family symmetry.
Though we also make use of quark-lepton symmetry and left-right symmetry, we do not utilize the family symmetry and assume that the Majonara masses have a hierarchical structure ("geometric mass hierarchy" \cite{Xing}) instead.
%%%%%%%%%%%%%%%%%%%%%%%%%%%%%%%%%%%%%%%%%%%%%%%%%%%%%%%%%%%%%%
\subsection{Inverted mass spectrum\label{imh}}
We can just follow the procedure given above for the normal mass spectrum, and find two values of $R$
\begin{eqnarray}
R
=
(2.927,6.027)\times 10^{-6}
=
(0.2462,0.5069)\times r^2.
\label{Rs}
\end{eqnarray}
The values of $\rho _1$ and $\rho _2$ take
\begin{eqnarray}
\rho _1
&\equiv&
\frac{m_1}{m_3}
=
(11.62,2.739),
\label{rho1Rs}\\
\rho _2
&\equiv&
\frac{m_2}{m_3}
=
(11.84,2.785),
\label{rho2Rs}
\end{eqnarray}
and we eventually get the absolute values of neutrino masses,
\begin{eqnarray}
m_1
&=&
(4.619,4.943)\times 10^{-2}~\textrm{eV}
\label{m1s}\\
m_2
&=&
(4.707,5.026)\times 10^{-2}~\textrm{eV}
\label{m2s}\\
m_3
&=&
(0.3976,1.805)\times 10^{-2}~\textrm{eV},
\label{m3s}
\end{eqnarray}
and the corresponding the greatest Majorana mass
\begin{eqnarray}
m_3^R
=
(14.70,3.861)\times 10^{15}~\textrm{GeV}.
\label{majoranamasss}
\end{eqnarray}

\begin{figure}[h]
\begin{center}
\includegraphics[scale=1.0]{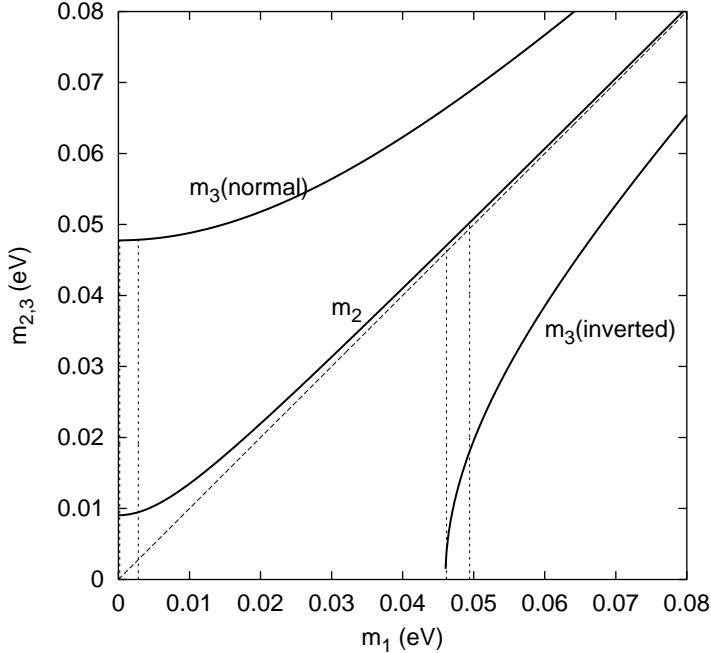}
\end{center}
\caption{
A schematic view of neutrino mass spectra to show the obtained results.
The horizontal coordinate corresponds to $m_1$ and the vertical coordinate corresponds to $m_2$ and $m_3$.
The solid curved lines show the constraints to mass-squared difference obtained by neutrino oscillation experiments and the crossing points with the vertical dotted lines show the obtained results in Eqs.(\ref{m1f}),(\ref{m2f}),(\ref{m3f}) and (\ref{m1s}),(\ref{m2s}),(\ref{m3s}).
}
\label{fig1}
\end{figure}
%%%%%%%%%%%%%%%%%%%%%%%%%%%%%%%%%%%%%%%%%%%%%%%%%%%%%%%%%%%%%%
\subsection{In the case of $\sin^2\theta _{13}<0.041$}
For simplicity, we have used the assumption $\theta _{13}=0$ in Eq.(\ref{t13rat}), as $\theta _{13}$ is stringently constrained.
It, however, will be desirable to evaluate the dependence of mass eigenvalues on $\theta _{13}$ at least for its 3$\sigma$ allowed range, i.e. $\sin^2\theta _{13}<0.041$.
There is no difficulty to achieve this, once we employ the same prescription as the one mentioned above.
We found that generally the solution for $m_{1,2,3}$ are multi-valued for each values of $\theta _{13}$.
At the same time, we found that the smallest mass eigenvalue is rather stable for the allowed $\theta _{13}$ in both cases (normal or inverted mass spectrum).
We thus get absolute lower bound on the smallest mass as follows.
\begin{itemize}
\item the case of normal mass spectrum:
\begin{eqnarray}
m_1\ge 1\times 10^{-4}\textrm{ eV}
\label{}
\end{eqnarray}
\item the case of inverted mass spectrum:
\begin{eqnarray}
m_3\ge 1\times 10^{-3}\textrm{ eV}
\label{}
\end{eqnarray}
\end{itemize}
The lower bound for the inverted  mass spectrum is very encouraging for the search for neutrinoless double beta decay \cite{Joaq}.
We also have found that for certain value of $\theta _{13}$ the Wilkinson microwave anisotropy probe (WMAP) results \cite{Bennett}, $\sum _im_i<0.70$ eV, is violated in both cases.

%%%%%%%%%%%%%%%%%%%%%%%%%%%%%%%%%%%%%%%%%%%%%%%%%%%%%%%%%%%%%
%%%%%%%%%%%%%%%%%%%%%%%%%%%%%%%%%%%%%%%%%%%%%%%%%%%%%%%%%%%%%%
%%%%%%%%%%%%%%%%%%%%%%%%%%%%%%%%%%%%%%%%%%%%%%%%%%%%%%%%%%%%%%
\section{RG evolution of Dirac neutrino mass matrix and reanalysis of the neutrino masses\label{rg}}
In previous sections, we derived the absolute values of neutrino masses, assuming $\mathcal{M}_D$ coincide with up-type quark mass matrix at the weak scale.
Such coincidence, however, is realized only at the GUT scale, and the relation between $\mathcal{M}_D$ and up-type quark mass matrix is modified by the renormalization group (RG) effect.
In this section, we thus improve the analysis of previous sections by including the RG effect.
In the computation of RG evolution of $\mathcal{M}_D$, we pay special attention to the successive decoupling of heavy right-handed neutrinos.

In the following analyses, the framework we work on is a supersymmetric $SO(10)$ GUT.
We use a chain of spontaneous gauge symmetry breaking in which $SO(10)$ GUT breaks into the standard model via the minimal supersymmetry standard model (MSSM) :
\begin{eqnarray}
SO(10)
\stackrel{\{16_H\}}{\rightarrow}
SU(5)
\stackrel{\{45_H\}}{\rightarrow}
SU(3)\times SU(2)\times U(1)
\stackrel{\{10_H\}}{\rightarrow}
SU(3)\times U(1)
,
\label{}
\end{eqnarray}
and for simplicity we assume that the symmetry breaking scales of $SO(10)$ and $SU(5)$ are sufficiently close to each other, i.e. $\langle 16_H\rangle\simeq\langle 45_H\rangle$ 
\footnote{
In the Yukawa sector, vev of heavy Higgs, $\langle\overline{126}_H\rangle=\langle(10,1,3)\rangle$ ((10,1,3) is a repr. of $SU(4)_{PS}\times SU(2)_L\times SU(2)_R$), gives a Majorana mass to $\nu _R$.
In addition, vevs of two light Higgs, $\langle 10_H\rangle=\langle(1,2,2)\rangle$ (this is a $SU(4)$ singlet), give the Dirac masses to quark and lepton sectors and we can achieve the desirable relations, $m_u=m_{\nu}$ and $m_d=m_e$ at the GUT scale \cite{Senja}.
}
 .
We need to introduce multiple number of 10-dimensional repr. of Higgs field, since otherwise the up-sector and down-sector of quarks and leptons have degenerate masses (at GUT scale).
Let us note that such problem is automatically evaded in our scheme, as supersymmetry inevitably demands two 10-repr. of Higgs fields.
The supersymmetry, of course, plays an important role in the unification of gauge couplings at the GUT scale.

At first, we evaluate the evolutions of Yukawa coupling matrices of up- and down-type quark sectors from the weak scale towards the GUT scale (up-stream) in MSSM scheme and identify the obtained Yukawa couplings at $M_{GUT}$ with those of neutrino and charged-lepton sectors at the GUT scale, respectively.
Then, we evaluate the evolutions of the Yukawa coupling matrices of lepton sectors from the GUT scale towards the weak scale (down-stream), taking into account the successive decoupling of three singlet neutrinos at their threshold energies.
Using the obtained $\mathcal{M}_D$ (the neutrino Yukawa coupling matrix $\times~v\cdot\sin\beta/\sqrt{2}$), we can immediately write down the effective neutrino mass matrix at low energy.
This mass matrix, however, still have three parameters, i.e. $\phi _{12},\phi _{23},\phi _{13}$, after we fix $R$ and $m_3^R$ to carry out the RG evolutions.
To fix these three parameters, we use the conditions for three mixing angles, Eqs.(\ref{t12rat}),(\ref{t23rat}),(\ref{t13rat}).
We then just follows the method we developed in the previous sections to get the eigenvalues of $\mathcal{M}_{\nu}$ and compare the obtained ratio of mass-squared differences with the observed one shown in Eq.(\ref{massrat}).

In MSSM, the RG equations of the Yukawa coupling matrices at one-loop are given by \cite{Cheng,Lindner,Lindner2}
\begin{eqnarray}
\frac{dY_a}{dt}
&=&
\frac{1}{16\pi^2}
\left(
\left(T_a-G_a\right){\bf 1}
+\sum _{b}C^b_a~Y_bY_b^{\dag}
\right)
Y_a,
\qquad(a=u,d,\nu,e)
\label{yukawarg}
\\
t&=&\ln\left(\mu/M_z\right),
\nonumber
\end{eqnarray}
\begin{eqnarray*}
T_u=T_{\nu}=3~Tr[Y_uY_u^{\dag}]+Tr[Y_{\nu}Y_{\nu}^{\dag}]
\qquad
T_d=T_e=3~Tr[Y_dY_d^{\dag}]+Tr[Y_eY_e^{\dag}]
\\
G_u=\frac{13}{15}g_1^2+3g_2^2+\frac{16}{3}g_3^2
\quad
G_d=\frac{7}{15}g_1^2+3g_2^2+\frac{16}{3}g_3^2
\quad
G_{\nu}=\frac{3}{5}g_1^2+3g_2^2
\quad
G_e=\frac{9}{5}g_1^2+3g_2^2
\\
C^u_u=C^d_d=C^e_e=+3,~C^{\nu}_{\nu}=+1
\qquad
C^u_d=C^d_u=C^{\nu}_e=C^e_{\nu}=+1
\qquad
\textrm{others}=0.
\end{eqnarray*}
Note that the coefficient $C_{\nu}^{\nu}$ we use is different from that used for Dirac neutrinos.
This is because in the seesaw mechanism the presence of the inverse right-handed Majorana mass matrix modifies the coefficient as $C_{\nu}^{\nu}=3\rightarrow 1$.
One problem here is that the evolutions of the Yukawa coupling matrices of quark and lepton sectors are coupled with each other since the term $T_a$, the contribution of Higgs self-energy diagram, contains both of quark and lepton Yukawa couplings.
It seems that when we solve the "up-stream" RG equation for quark Yukawa couplings, we need to know the values of leptonic Yukawa couplings, which are unknown.
Fortunately, it turns out that in the estimation of $\mathcal{M}_D$ at the weak scale we can safely neglect the interference from another sector, as we show below.
Namely, the Eq.(\ref{yukawarg}) formally can be solved as
\begin{eqnarray}
Y_a(t)
=
\exp
\left[
\frac{1}{16\pi^2}
\int^t_0dt~
\left(
\left(T_a-G_a\right){\bf 1}
+\sum _{b}C^b_a~Y_bY_b^{\dag}
\right)
\right]
Y_a(0).
\label{}
\end{eqnarray}
Since the exponent is of $\mathcal{O}(\alpha)$, we can expand this solution perturbatively up to its leading-order as
\begin{eqnarray*}
Y_a(t)
\approx
\left[
1
+
\frac{1}{16\pi^2}
\int^t_0dt~
\left(
\left(T_a-G_a\right){\bf 1}
+\sum _{b}C^b_a~Y_bY_b^{\dag}
\right)
\right]
Y_a(0).
\end{eqnarray*}
Applying this arguments to the Yukawa coupling matrices of up-type quark and neutrino sector, and identifying these matrices at the GUT scale ($\mu\simeq 2\times 10^{16}$ GeV) , i.e. $Y_u(t_{GUT})=Y_{\nu}(t_{GUT})$, $Y_{\nu}(0)$, of our interest, can be expressed as
\begin{eqnarray*}
Y_{\nu}(0)
&=&
\left[
1
-
\frac{1}{16\pi^2}
\int^{t_{GUT}}_0dt~
\left(
\left(T_{\nu}-G_{\nu}\right){\bf 1}
+\sum _{b}C^b_{\nu}~Y_bY_b^{\dag}
\right)
\right]
Y_{\nu}(t_{GUT})
\\
&=&
\left[
1
-
\frac{1}{16\pi^2}
\int^{t_{GUT}}_0dt~
\left(
\left(T_{\nu}-G_{\nu}\right){\bf 1}
+\sum _{b}C^b_{\nu}~Y_bY_b^{\dag}
\right)
\right]
\\
&&
\times
\left[
1
+
\frac{1}{16\pi^2}
\int^{t_{GUT}}_0dt~
\left(
\left(T_u-G_u\right){\bf 1}
+\sum _{b}C^b_u~Y_bY_b^{\dag}
\right)
\right]
Y_u(0)
\\
&\simeq&
\left[
1
+
\frac{1}{16\pi^2}
\int^{t_{GUT}}_0dt~
\left(
\left(G_{\nu}-G_u\right){\bf 1}
+\sum _{b}(C^b_u-C^b_{\nu})~Y_bY_b^{\dag}
\right)
\right]
Y_u(0),
\end{eqnarray*}
where we have used the relation $T_{\nu}=T_u$.
Therefore, the problematic factor $T_a$ has disappeared, and we can treat the RG equations for quark sector and lepton sector independently.

Adopting the input data set of the Yukawa coupling matrices of quark sectors at the weak scale ($\mu\simeq M_Z$) \cite{Koide}:
\begin{eqnarray*}
m_i=diag(Y_u)_{ii}\times \frac{v\cdot \sin\beta}{\sqrt{2}}\qquad(i=u,c,t)
\\
m_j=diag(Y_d)_{jj}\times \frac{v\cdot \cos\beta}{\sqrt{2}}\qquad(j=d,s,b)
\\
m_u=2.33\textrm{ MeV}
\qquad
m_c=677\textrm{ MeV}
\qquad
m_t=181\textrm{ GeV}
\\
m_d=4.69\textrm{ MeV}
\qquad
m_s=93.4\textrm{ MeV}
\qquad
m_b=3\textrm{ GeV}
\\
\sin\theta _{12}=0.2205
\qquad
\sin\theta _{23}=0.0373
\qquad
\sin\theta _{13}=0.003
\end{eqnarray*}
, we can carry out solving the up- and down-stream RG equations.

In order to know the threshold energies of decoupled singlet neutrinos, we need to know the values of $R$ and $m_3^R$.
We scan the following ranges of these parameters,
\begin{eqnarray}
1\times 10^{-6}<R<1\times 10^{-3}
\label{}
\\
1\times 10^{13}<m_3^R<2\times 10^{16}\textrm{ (GeV)},
\end{eqnarray}
and set $\tan\beta=10$ as a reference value (in fact, the obtained values are stable for $\tan\beta=1,10$ and $30$).
There are no difficulty in solving the up-stream RG equation.
In the case of down-stream RG equation, however, we have to take into account the decoupling effects of singlet neutrinos at their threshold energies $m_3^R,m_3^R\cdot R$ and $m_3^R\cdot R^2$.

To clarify our prescription, we write the neutrino Yukawa coupling matrix as
\begin{eqnarray}
\mathcal{L}_{yukawa}
=
-\sum _{i,j=1}^3
\overline{\nu _{iL}}(Y_{\nu})_{ij}\nu _{jR}\phi^0.
\label{}
\end{eqnarray}
In the energy region, $m_3^R<\mu$, the evolution of $Y_{\nu}$ is calculated by use of full members of $Y_{\nu}$.
Once $\mu$ passes through $m_3^R$, i.e. in the region of $m_3^R\cdot R<\mu<m_3^R$, the evolutions of $Y_{\nu}$ is calculated by use of the subset of $Y_{\nu}$ obtained by removing 3rd column of $Y_{\nu}$ in the product of $Y_{\nu}$, i.e. $(Y_{\nu}Y_{\nu}^{\dag})_{ij}=(Y_{\nu})_{i1}(Y_{\nu}^{\ast})_{j1}+(Y_{\nu})_{i2}(Y_{\nu}^{\ast})_{j2}$, since $\nu _{3R}$ is decoupled in this region.
We can repeat the same procedure at each step of decoupling of single neutrinos, and we use the obtained $Y_{\nu}$ at $M_{W}$ to calculate $\mathcal{M}_{\nu}$.

\begin{figure}[h]
\begin{center}
\includegraphics[scale=1.0]{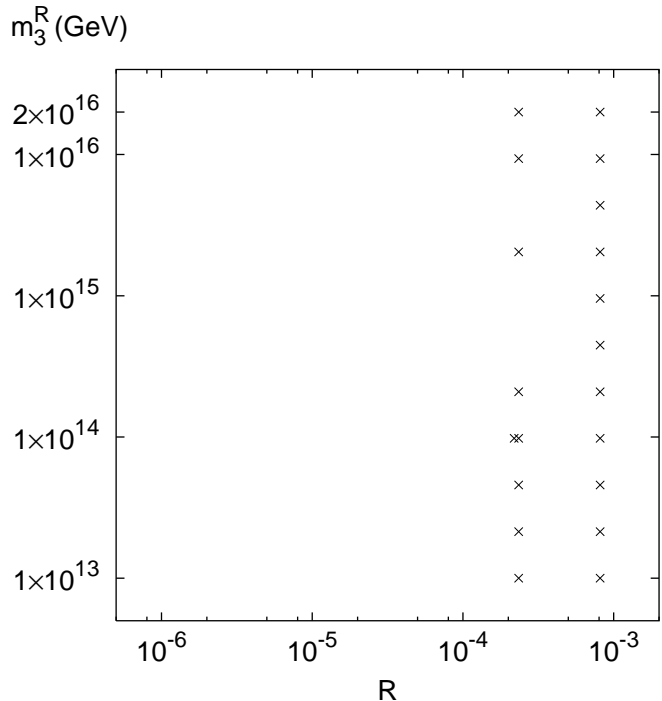}
\end{center}
\caption{
The horizontal coordinate corresponds to $R$ and the vertical one corresponds to $m_3^R$.
At the lattice points marked by cross, the obtained values of $\Delta=\frac{m_2^2-m_1^2}{m_3^2-m_2^2}$ is compatible with the allowed range $7.9/280<\Delta<8.5/180$ ($3\sigma$ C.L.) coming from neutrino oscillation experiments.
}
\label{figRM}
\end{figure}

\begin{figure}[h]
\begin{center}
\includegraphics[scale=1.0]{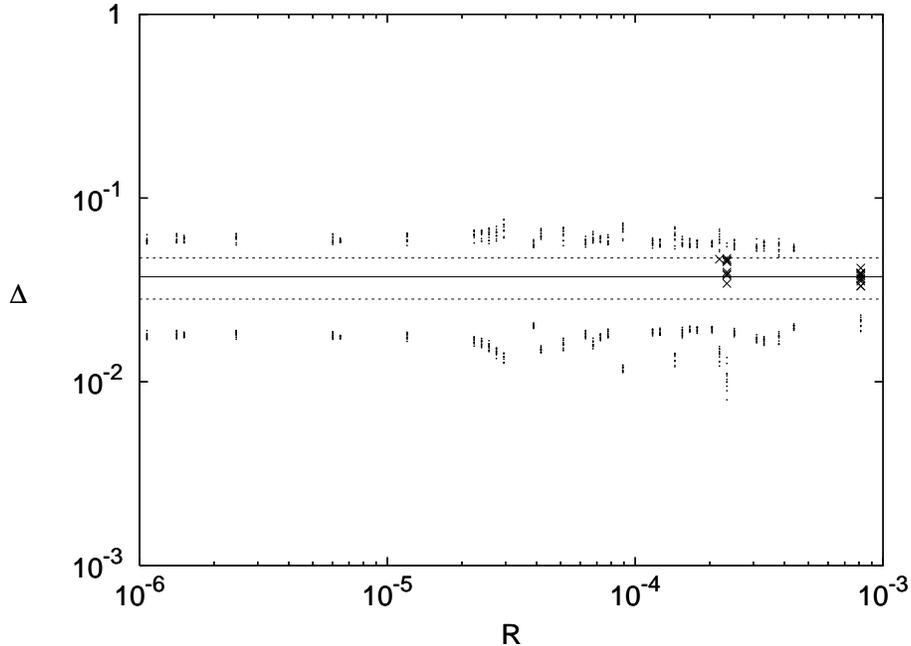}
\end{center}
\caption{
The obtained values of $\Delta=\frac{m_2^2-m_1^2}{m_3^2-m_2^2}$ as the function of $R$.
Multiple points for each $R$ correspond to different values of $m_3^R$.
We have taken $\tan\beta=10$.
The horizontal solid line corresponds to the best fit value of $\Delta$ from experiments, $\Delta=8.2/220$, and the horizontal dotted lines show the allowed range of $\Delta$ at $3\sigma$ C.L., i.e. $7.9/280<\Delta<8.5/180$.
}
\label{figRDelta}
\end{figure}

\begin{figure}[h]
\begin{center}
\includegraphics[scale=1.0]{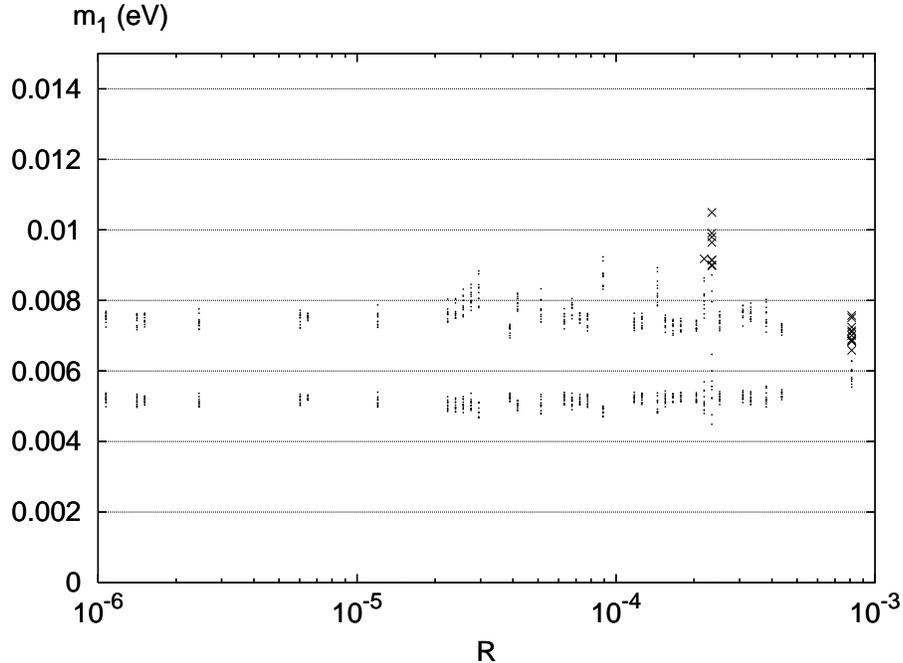}
\end{center}
\caption{
The obtained values of $m_1$ as the function of $R$.
The horizontal coordinate corresponds to $R$ and the vertical coordinate corresponds to $m_1$ (eV).
The points marked by cross satisfy the condition, $7.9/280<\Delta<8.5/180$ ($3\sigma$ C.L.), and the other do not satisfy this condition.
}
\label{figRmass}
\end{figure}

The obtained results are shown in Fig.\ref{figRM} and Fig.\ref{figRDelta}, where we define $\Delta=\frac{m_2^2-m_1^2}{m_3^2-m_1^2}$.
All of them satisfy the condition that $m_2<m_3$, i.e. normal mass spectrum.
According to these figures, $\Delta$ is rather insensitive to the values of $m_3^R$.
On the other hand, it is interesting to know that the allowed region for $R$ is restricted to the range around $1\times 10^{-4}$.
Since we fix the effective neutrino mass matrix at low energy completely, we can also analyze the absolute values of neutrino masses simultaneously and the obtained results are shown in Fig.\ref{figRmass}.
We can find in this figure that the absolute values of lightest neutrino mass are stable significantly around certain region, $5\times 10^{-3}<m_1<1\times 10^{-2}$ (eV).
Note that these are higher than what we obtained in previous section (Eq.(\ref{m1f})).

%%%%%%%%%%%%%%%%%%%%%%%%%%%%%%%%%%%%%%%%%%%%%%%%%%%%%%%%%%%%%%
%%%%%%%%%%%%%%%%%%%%%%%%%%%%%%%%%%%%%%%%%%%%%%%%%%%%%%%%%%%%%%
\section{Conclusion\label{concl}}
In this paper, we derive the absolute values of three neutrino masses only invoking to the seesaw mechanism collaborated by the unification of neutrino Dirac mass matrix with that of up-type quarks.
Especially for $\theta _{13}=0$, the obtained results are $m_1:m_2:m_3\approx 1:3:17$ or $1:50:250$ in the case of normal mass spectrum of neutrino masses and $m_1\simeq m_2:m_3\approx 2:1$ or $12:1$ in the case of inverted mass spectrum and we found that there exists the lower bound of the smallest neutrino mass for the allowed range of $\theta _{13}$ at 3$\sigma$ C.L.
In addition, we reanalyze the neutrino Yukawa coupling matrix at low energy including the effect of the RG evolution, paying a special attention to the successive decoupling effects of three singlet neutrinos and derive the ratio of mass-squared differences applying a method derived in previous sections.
In this analyses, we find that in certain region of $R$ there exist the solutions which are compatible with the experimental data of neutrino oscillations and that the absolute values of neutrino masses are higher that what we obtained in previous sections.

In the process, we impose some suitable assumptions for simplicity because our main purpose is to clarify the procedure to derive the neutrino masses, where large mixing angles observed in the leptonic sector is achieved invoking to the unification inspired by GUTs.
Let us recall that, for instance, in $SO(10)$ GUT, which incorporate left-right symmetry $SU(4)_{PS}\times SU(2)_L\times SU(2)_R$, the neutrino Dirac mass matrix is naturally unified with that of up-type quarks and they are identical at least at the scale of $M_{GUT}$.

However, there still remain some questions about the justification of these assumptions.
For example, it is an interesting question to ask, though difficult to answer, how the obtained results are modified when we take into account the 6 CP violation phases which in general embodied in Majorana mass matrix as physical degrees of freedom resulting from seesaw mechanism.

These remaining questions will be addressed in ref.\cite{pre}.
%%%%%%%%%%%%%%%%%%%%%%%%%%%%%%%%%%%%%%%%%%%%%%%%%%%%%%%%%%%%%%
%%%%%%%%%%%%%%%%%%%%%%%%%%%%%%%%%%%%%%%%%%%%%%%%%%%%%%%%%%%%%%
%\appendix
% If you have acknowledgments, this puts in the proper section head.
\begin{acknowledgments}
I thank C. S. Lim for helpful and fruitful discussions and for careful reading and correcting of the manuscript.
I also thank D. Kitagawa for helpful discussions on numerical and analytic calculations.
\end{acknowledgments}
%%%%%%%%%%%%%%%%%%%%%%%%%%%%%%%%%%%%%%%%%%%%%%%%%%%%%%%%%%%%%%
%%%%%%%%%%%%%%%%%%%%%%%%%%%%%%%%%%%%%%%%%%%%%%%%%%%%%%%%%%%%%%

%%%%%%%%%%%%%%%%%%%%%%%%%%%%%%%%%%%%%%%%%%%%%%%%%%%%%%%%%%%%%%
%%%%%%%%%%%%%%%%%%%%%%%%%%%%%%%%%%%%%%%%%%%%%%%%%%%%%%%%%%%%%%
%%%%%%%%%%%%%%%%%%%%%%%%%%%%%%%%%%%%%%%%%%%%%%%%%%%%%%%%%%%%%%
%%%%%%%%%%%%%%%%%%%%%%%%%%%%%%%%%%%%%%%%%%%%%%%%%%%%%%%%%%%%%%
\end{document}